\documentclass[british, 12pt, a4paper]{article}
\usepackage[left=1in,top=1in,right=1in,bottom=1in,nohead,paperwidth=8.5in, paperheight=11in]{geometry} 
\usepackage{amsmath}
\usepackage{stmaryrd}
\usepackage{natbib} 
\usepackage{float}
\usepackage{rotating}
\usepackage{multirow}
\usepackage{algorithm}
\usepackage{todonotes}
\usepackage[titletoc, title]{appendix}
\usepackage{adjustbox}
\usepackage{graphicx}
\usepackage{bbm}
\usepackage{xcolor}
\usepackage{subcaption}
\usepackage{caption}
\usepackage{setspace}
\usepackage{soul}
\usepackage[normalem]{ulem}
\usepackage{hyperref}

\graphicspath{{./Figures/}}

\vspace{-2cm}

\title{\bf 
Sentiment Analysis of Economic Text:\\A Lexicon-Based Approach
}

\author{
 Luca Barbaglia\thanks{
 \thanks{}European Commission, Joint Research Centre, {\tt e-mail}: luca.barbaglia@ec.europa.eu (corresponding)
 }, 
 Sergio Consoli\thanks{European Commission, Joint Research Centre, {\tt e-mail}: sergio.consoli@ec.europa.eu } 
 Sebastiano Manzan\thanks{Baruch College, United States, {\tt e-mail}: sebastiano.manzan@baruch.cuny.edu}, \\
 Luca Tiozzo Pezzoli\thanks{University of the Balearic Islands, Spain and Brunel University, United Kingdom, {\tt e-mail}: luca.tiozzo-pezzoli@uib.es}, and 
 Elisa Tosetti\thanks{University of Padua, Italy, {\tt e-mail}: elisa.tosetti@unipd.it}
 }

\vspace{-2cm}
 
\date{}

\begin{document}
\maketitle 
\thispagestyle{empty}


\begin{abstract}
\begin{center}
\mbox{}\\
\begin{minipage}{13cm}
\noindent\linespread{1}\selectfont
{\small
We propose an Economic Lexicon (EL) specifically designed for textual applications in economics. We construct the dictionary with two important characteristics: 1) to have a wide coverage of terms used in documents discussing economic concepts, and 2) to provide a human-annotated sentiment score in the range [-1,1]. 
We illustrate the use of the EL in the context of a simple sentiment measure and consider several applications in economics. The comparison to other lexicons shows that the EL is superior due to its wider coverage of domain relevant terms and its more accurate categorization of the word sentiment.
}
\mbox{}\\
\mbox{}
{\footnotesize {\it Abbreviations}: 
ADS, Aruoba and Diebold and Scotti;
AUC, Area Under the Curve;
CBOE, Chicago Board Options Exchange;
EL, Economic Lexicon;
EP, Economic Pessimism;
EPU, Economic Policy Uncertainty;
FEARS, Financial and Economic Attitudes Revealed by Search;
FOMC, Federal Open Market Committee;
FSS, Financial Stability Sentiment;
LMD, Loughran and McDonald;
MCSI, Michigan Consumer Sentiment Index;
NBER, National Bureau of Economic Research;
REN, Renault;
SSW, Shapiro and Sudhof and Wilson;
VADER, Valence  Aware  Dictionary  for sEntiment Reasoning;
VIX, Volatility Index;
WBG, World Bank Group.
}\\
\end{minipage}
\end{center}
\end{abstract}


\newpage
\pagenumbering{arabic}

\section{Introduction}


Textual data such as news, policy statements, and transcript of speeches are increasingly used in economics and finance \citep{gentzkow2019text,algaba2020econometrics}. 
Some recent applications use text to measure uncertainty \citep{Bloom_Baker_Scott_2016}, evaluate the effect of central bank communication \citep{hansen2018transparency, ter2021narrative}, and to gauge economic sentiment (\citealp{balke2002well,Tetlock_2007, armesto2009measuring, Garcia2013, Nardo2016, Thorsrud_2018, shapiro2020measuring, barbaglia2020forecasting}). 
In these cases, text is used to proxy for an unobservable variable of interest (e.g., uncertainty or sentiment) or to evaluate the effect that news, speeches, and policy-maker statements have on macroeconomic outcomes. 
In several of these applications the objective is to {\color{black} estimate the sentiment embedded in the text, that is,} to determine whether the reader is likely to interpret the text as providing a positive or negative perspective regarding a topic \citep{pang2008opinion}. 

The goal of this paper is to develop a domain-specific Economic Lexicon (EL) that can be used to measure sentiment in a more accurate way relative to existing lexicons. {\color{black}The objective of EL is to be useful in many different contexts within economics rather than being tailored to a specific economic application (e.g., monetary policy documents). }
The two main challenges in constructing a lexicon are (i) the selection of a large set of words that are likely to convey sentiment in economic applications, and (ii) to quantify the word's tone. \\
To address the first challenge, we do an extensive work of word selection based on a comprehensive corpus that includes more than 13 million news articles for the major US and UK journals, as well as 469 documents produced by the Federal Reserve and the European Central Bank, for a total of over 7.5 billion words. 
We use this textual data set to extract sentences that include a wide range of economic concepts such as employment, layoffs, industrial production, and house prices, among many others. 
Within each sentence, we examine the dependence structure to identify the adverbs, verbs, and adjectives that are used as modifiers of the meaning of the economic concept. 
Hence, our strategy to select the relevant terms to include in the EL relies on two refinements: first, we isolate the sentences that contain an economic keyword, and then we extract the words that are semantically related to the economic concept. 
We believe that this approach is better able to separate the words that are uniquely used and relevant in an economic context. 
Based on this selection of words, we then analyze their frequency to select the most common words to include in our dictionary.\\
To address the second challenge, we measure the sentiment elicited by each word in an economic context.

A key contribution of this paper is that we ask human annotators to provide a sentiment score for each of the over 7 thousand words included in the EL.
The human annotation is an approach that is also followed by \cite{Loughran_McDonald_2011} (which we denote LMD henceforth) for applications in finance and accounting and, more recently, by \cite{correa2021sentiment} that develop a dictionary for financial stability. There are also several proposals of economic lexicons that rely on a model to assign sentiment to the words {\color{black} \citep{renault2017intraday, GARDNER2021, shapiro2020measuring}}. 
More specifically, we ask 10 {\color{black} human} annotators to provide a score in the interval between -1 and 1 that quantifies the subjective sentiment of a word in the context of an economic sentence. 
This approach is similar to \cite{Loughran_McDonald_2011} in which the authors classified the terms in their data set \textcolor{black}{in various categories.}
Once we have elicited the tone, we take the median score across the annotators to represent the sentiment associated with the word. 
The fine grid of values between -1 and 1 provides an assessment of the strength of the sentiment that is elicited by a term, as opposed to considering all positive (negative) words equally meaningful. 
On a more practical note, a fine grid of sentiment scores can always be converted to a classification of the terms in positive and negative, as in LMD, if requested by the application.

We then compare the proposed EL against LMD and the dictionaries proposed by \cite{renault2017intraday} and \cite{shapiro2020measuring} (which we denote REN and SSW, respectively) in the context of some relevant economic applications.
An important feature of the most recent lexicons is that they significantly expand the set of terms, in particular regarding words with positive sentiment. 
When considering the sentiment assigned to each term, we find that there is positive correlation in the score of the terms that are in common across dictionaries, despite some substantial disagreement regarding the sentiment of other terms. 
Since the work of \cite{Tetlock_2007}, the dominant view is that sentiment in economic and financial applications is mostly generated by negative words. 
However, \cite{Garcia2013} uses the LMD dictionary and finds that also positive terms contribute to predict daily returns. 
Considering an extensive sample of economic news, we find that words with a positive sentiment occur often in our sample and with a frequency similar to negative words. 
In addition, our results suggest that the sentiment deriving from negative words is more variable and pro-cyclical relative to the sentiment from positive terms. 
This probably explains the earlier findings that pessimism is overwhelmingly due to negative words. 


We compare the performance of the lexicons in the context of a measure of Economic Pessimism (EP) that is calculated as the difference between the fraction of negative and positive terms in a document. 
The EP measure is thus lexicon-specific and might diverge because dictionaries include different terms, and also due to the diverse sentiment classification of each word. 
We find that the EP is correlated with indicators of economic and financial uncertainty, consumer sentiment, and has predictive power to forecast recessions. 
In particular, we find that the EP measure computed using the EL is significant and, once included, makes the other EP measures irrelevant. 
We show that the superiority of the EL derives from the inclusion of more economic terms relative to the other dictionaries, and also because of a more accurate classification of the term sentiment.

The remainder of the paper is structured as follows.
Section \ref{sec:literature} reviews the most important works on lexicon-based approaches for economic and financial analysis.
In Section \ref{data},  we describe the data sources used for constructing the dictionary. 
Section \ref{dictionary} outlines the steps followed for designing our lexicon and the scores, while Sections \ref{application} and \ref{econ_applications} illustrate the use of the proposed dictionary in several empirical applications. 
Finally, Section \ref{conclusions} concludes.

\section{Literature review}
\label{sec:literature}

Several studies that aim at measuring sentiment in economics and financial documents typically search for specific terms within a document. 
An influential example is provided by \cite{Bloom_Baker_Scott_2016} that constructs an indicator of policy-related economic uncertainty, known as EPU, by searching for a set of relevant keywords in newspaper articles.
The EPU index is used extensively as a proxy for uncertainty in various economic and financial applications (see \citealp{Bernal_Gnabo_Guilmin_2016}, and \citealp{altig2020economic}, among others). Although the EPU index, per se, is not a sentiment indicator based on a lexicon, it {\color{black} can be} interpreted as one since it measures the frequency of the word ``uncertainty'' that has a specific sentiment connotation in the economic domain. 
Another instance of a keyword-based index is provided by \cite{Engelberg2015}. They measure market sentiment by the FEARS (Financial and Economic Attitudes Revealed by Search) index that is based on Internet search queries of the terms ``recession'', ``unemployment'' and ``bankruptcy''. The goal of this index is to capture the negative mood of investors during recessions. 
Also in this case sentiment is not measured directly but is embedded in the choice of terms that are used to select the articles, such as ``uncertainty" and ``recession", that have a clear connotation in economics. 

A number of studies use the Harvard IV-4 dictionary to measure sentiment from economic and financial text. \cite{Tetlock_Tsechansky_Macskassy_2008} show that the fraction of negative words from the Harvard IV-4 negative word list found in firm-specific news stories helps forecasting stock prices. 
\cite{Loughran_McDonald_2011} argue that the Harvard IV-4 word list might not be suitable to measure sentiment in financial and accounting documents. The reason is that many terms that are classified as positive or negative do not have such a connotation in the financial context. In an analysis of a large sample of annual 10-K company reports they also find that almost three-fourths of the words identified as negative by the Harvard IV-4 were typically not considered negative in a financial context. For example, terms like \textit{depreciation} or \textit{liability} do not have polarity when occurring in a 10-K report. Based on textual data from the reports, they select a set of 4,140 terms that they consider relevant for sentiment in a business and financial context and then classify these words into six sentiment categories: negative, positive, uncertainty, litigious, strong and weak modal. The negative and positive sentiment categories include 2,355 and 354 words, respectively. 
This lexicon has been adopted by several studies to determine sentiment in economic and financial applications (see, among others, \citealp{Garcia2013}, \citealp{hansen2016shocking}, \citealp{hassan2019firm}, \citealp{shapiro2022taking}). 

The lexicon proposed by \cite{Loughran_McDonald_2011} is domain-specific, which allows researchers to obtain more accurate measures of sentiment relative to the case of general dictionaries. However, several papers have found that the LMD lexicon might not be suitable for applications in economics. 
One important drawback is that LMD contains many terms related to finance and accounting that rarely appear in documents discussing economics. For instance, words such as \textit{alienation, condoned, juris, obligor, pledgor} are very rarely used in an economic context and, even when they do, their inclusion may capture marginal aspects of the economic topic being discussed. 
More specifically, we calculate the frequency of the terms included in LMD on 13 million UK and US newspaper articles published over 40 years. We find that 2,057 terms (about 50\% of the dictionary) never appear in the articles, while 615 terms occur 10 times or less.
At the same time, we observe that sentiment-rich terms that are often used by journalists to comment on the economic situation do not appear in the LMD list. 
Notable examples are the words \textit{fall} or \textit{rise} that occur over 56,000 and 54,000 times, respectively, in our sample of economic news. 
For these reasons, the LMD lexicon might not be a reliable lexicon to measure sentiment in economic documents.

There have been several other attempts to develop lexicons that are designed for economic applications. \cite{correa2021sentiment} propose a dictionary with the specific purpose of measuring sentiment in the financial stability reports of Central Banks. Specifically, the authors use their proposed lexicon to construct a Financial Stability Sentiment (FSS) index and find that it captures changes in financial cycle indicators related to credit, asset prices, and systemic risk. \cite{GARDNER2021} develop a lexicon based on the statements of the Federal Open Market Committee (FOMC) for the 2000-2019 period. They then use the lexicon to compute a sentiment indicator for five topics, namely labor, output, inflation, financial, and future monetary policy. The paper evaluates the predictive ability of their sentiment index to forecast monetary policy decisions and macroeconomic outcomes and finds that it is an important predictor of FOMC decisions. 
\cite{shapiro2020measuring} construct an extended lexicon that includes terms from several existing lexicons. They then assign a sentiment score to each word based on the frequency of co-occurrence of the word and the sentiment categories (positive, neutral, and negative) of the sentences obtained using the Vader algorithm proposed by \cite{hutto2014vader}. The sentiment score of a word is thus positive if the word is more likely to occur in sentences that have a positive, relative to negative, VADER score. In this case, the sentiment of each word is model-based rather than being the result of human annotation as for the EL. Furthermore, the paper compares the ability of several lexicons to accurately measure the sentiment of 800 articles that have been categorized by human annotators as negative, neutral, and positive. The results show that the proposed lexicon performs better relative to several alternative lexicons and their combination, although the improvement is quite small.

\section{Data sources\label{data}}

To construct the EL we use different sources of documents discussing economic concepts. One source is represented by all articles retrieved from the Dow Jones Factiva platform for the six leading newspapers in the United States and the United Kingdom\footnote{The US newspapers are The New York Times, The Wall Street Journal, The Washington Post, The Dallas Morning News, The San Francisco Chronicle, and The Chicago Sun-Times, while for the UK we consider The Times, The Guardian, Daily Mail, The Economist, Evening Standard, The Sunday Times, and the Observer.}.
In particular, we collect articles published between January 1st, 1980 and December 31st, 2020. The data set provides information on the date of publication of the article, its title, body and a categorisation of the article's topic according to a classification provided by Factiva. 
We use the Factiva classification to discard stories related to sports which leaves us with approximately 13 million articles.
An additional source of economic text is represented by the full text of all Beige Book published by the Federal Reserve between 1983 and 2020, and the Economic Bulletin released monthly by the European Central Bank starting in 1999. A summary of the sources of information together with summary statistics is provided in the online Appendix. By combining information from journal articles and central bank documents we create a diverse textual data set that can provide a broad set of terms to include in our lexicon. 

\section{Lexicon design\label{dictionary}}

We follow four steps in designing the EL. 
The first is to identify the sentences in each document that discuss topics that are relevant from an economic perspective. This allows us to analyze locally the words that are used in combination with an economic concept, rather than considering the whole document that might contain sentences irrelevant to the context of interest. 
The second step of our analysis is to identify the words that are used to discuss the economic concept in the sentences selected earlier. More specifically, we parse the grammatical dependence of each term and select only the words that have a direct relationship with the economic concept included in the sentence. 
This provides us with a list of words that constitutes our lexicon. 
The following step consists of annotating the list of words with a sentiment score in the interval between -1 and 1. 
The analysis of the scores across human annotators shows that some words are interpreted very differently, although there is significant consensus among them for the vast majority of the words. In our EL we include only words with no or negligible ambiguity on the positive or negative connotation of the word.
In the rest of this section, we discuss in more detail the methodology that we follow in the development of the lexicon\footnote{The EL is available for download at \url{https://data.jrc.ec.europa.eu/dataset/1c054ef4-561a-464a-9077-3f6b09630da2}}.

\subsection{Economic concepts}
\label{sec:econ_concepts}
To determine the sentences in an article that are relevant to an economic context,
we compile a list of terms that contains concepts, categories, or more generally, entities belonging to the economic domain. 
To this end, we rely on two sources. 
The first is the World Bank Group's (WBG) Topical Taxonomy\footnote{WBG Topical Taxonomy available at \url{https://vocabulary.worldbank.org/taxonomy.html}.} that represents {\color{black} an open-source} classification in topics of the World Bank's knowledge and activities\footnote{For example, the topic {\it Economic Growth} has several sub-topics and among them {\it Determinants of Economic Growth}, which has itself several sub-topics such as {\it Macroeconomic Stability and Growth} which represents the terminal node with sub-topics {\it Fiscal Policy and Growth} and {\it Monetary Policy and Growth}.}. Examples of unigrams extracted from the WBG taxonomy are \textit{unemployment} and \textit{trade}, while examples of bigrams are \textit{economic growth}, \textit{macroprudential policy} and \textit{financial sector}. 
{\color{black} The benefit of using the WBG Taxonomy is that it provides a structured list of 3,888 concepts covering a wide range of terms that are used to describe economic activity, at least in relation to the WBG work. However, such taxonomy includes also terms that might not be closely related to economics. We thus decided to exclude those terms since they might have led to the selection of text unrelated to our purpose} (e.g., we excluded the term ``security''). {\color{black} We also complement the concepts related to the WBG activities, with those based on the central bank activities that are very often discussed in newspapers.}
{\color{black} For this reason, we also analyze} the Beige Book of the Federal Reserve and the monthly Economic Bulletin of the European Central Bank. The goal of these documents is to provide a periodic assessment of the economic conditions at both the regional and national levels. 
We consider these documents a relevant textual source to identify the entities that are often used to discuss the state of the economy {\color{black} and monetary policy}. 

By combining terms obtained from the WBG taxonomy and the central bank documents we collect a list of 355 entities (available in the online Appendix) that represents the basis for the construction of our lexicon. Based on this list, we then identify the sentences in our textual data that contain one or more of these terms and discard all other sentences as irrelevant to the economic context. 


\subsection{Dependency parsing}

\begin{figure}
 \centering
 \includegraphics[scale=0.45]{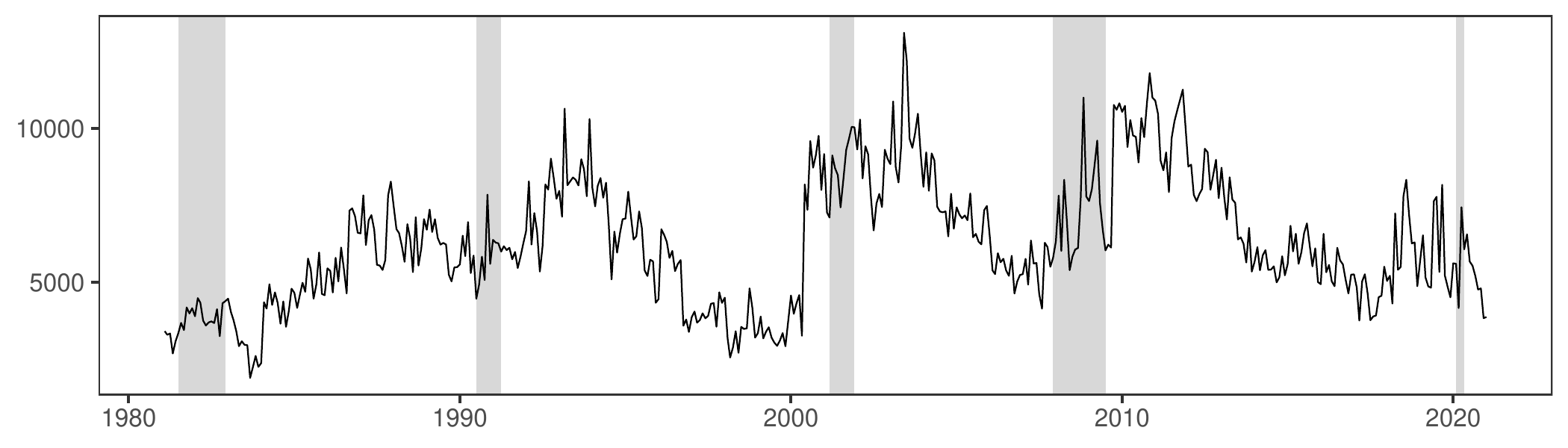}
 \\
 \caption{Time series of the number of noun phrases extracted aggregated at the monthly frequency. The grey areas correspond to the recessions as indicated by National Bureau of Economic Research (NBER) for the US.}
 \label{fig:chunks}
\end{figure}

The objective of dependency parsing is to understand the syntactic structure of a sentence by analysing the direct binary relations among pair of words\footnote{ We follow the classification proposed by \cite{de-marneffe-etal-2014-universal} that is composed of a set of 42 different types of ``universal" grammatical relations that are common to most languages.}. 
This allows us to identify the so called {\it noun phrases} that represent phrases that have a noun, single or compound, as the head and a set of dependent words that characterize its meaning. 
These terms are referred to as modifiers and consist of verbs, adverbs, and adjectives, among others. 
An example is provided by the phrase ``The economy suffered a slowdown", where the word ``economy" represents the subject while ``suffered" and ``slowdown" are modifier words. 
Once we have identified all the noun phrases, we extract those in which the core subject is in the list of economic concepts discussed earlier. This leads to the selection of more than 3 million noun phrases with the monthly frequency shown in Figure \ref{fig:chunks}. Interestingly, we observe {\color{black} a rapid increase in the number of sentences discussing economic concepts at the beginning of recessionary periods and a slow decline after their ends, which is similar to the behavior of the unemployment rate during and following slowdowns. It is possible that when the economy is slowing down there is more attention by newspapers about the economic situation, the expected depth and length of the recession, and the monetary and fiscal policies put in place to smooth the business cycle. Once the economy starts recovering, the focus shifts to how quickly and how strongly the economy will rebound, which typically takes longer to develop across sectors.} In terms of the construction of the lexicon, this dynamics might lead to oversampling terms with a negative sentiment while undersampling terms that convey a positive tone. We will investigate this issue further later in the paper.

The dependency parsing of the noun phrases produces a list of 84,019 modifier words. Since many of these terms appear very infrequently, we decided to narrow down the list to only words that occur at least 65 times. This reduces the list to approximately 12,000 terms\footnote{The choice of the threshold is mostly dictated by the budget available to reward the human annotators. A similar value for the threshold was also employed by \cite{renault2017intraday}.}. 
The analysis of the reduced list indicates that some words convey a sentiment, such as ``fall", ``decline", or ``loss", while other terms, like ``undertaking", ``supplementary", or ``above", might be considered neutral. 
Since the human annotation of each word is costly, we decided to further reduce the word list to terms that are more likely to be relevant for sentiment analysis. More specifically, each term was evaluated by four distinct human annotators (among the five authors of this paper) and we decided to include in our lexicon only terms that are considered by at least two of the four annotators as conveying sentiment. 

Our lexicon is composed of the list of modifier words and the economic entities described earlier. We include in the lexicon, and thus in the annotation exercise, the list of economic entities in order to assign a sentiment score also to these terms. This is relevant for sentiment analysis since modifiers that have a positive sentiment (e.g., {\it high}) of a negative economic concept (e.g., {\it unemployment}) acquire the connotation of the concept. 
Having a sentiment score also for the concept is useful, in particular, if a researcher intends to calculate a fine-grained sentiment score of the available text \citep{barbaglia2020forecasting,barbaglia2021forecasting}. 
The final list of words that constitutes the EL includes 7,295 terms. The only remaining step in the construction of the EL is the annotation that delivers a sentiment score for each word.

\subsection{Human annotation\label{sec_annotation}}

In order to attribute a tone to each word in the EL, we ask 10 human annotators {\color{black} representative of the US population} how they perceive its sentiment in the context of a phrase with economic content. 
In particular, we collect annotations from a US-based company specialised in annotation tasks\footnote{\color{black} We evaluated the performance of several companies by comparing the annotations on a sample of words. More specifically, we compared the agreement of the annotations provided by different companies with those provided by the authors of this paper.  We found that the annotations {\color{black} broadly} consistent with our judgment were provided by iMerit, a US-based company specialized in data labeling and annotation (\url{https://imerit.net}).}. To improve and facilitate the task, we provided annotators with a phrase for each term that shows the typical use of the word in an economic context. 
We select phrases that do not include a negation, as well as phrases with economic entities \textcolor{black}{as their head} that we believe can be interpreted negatively. 
For instance, we avoid sentences \textcolor{black}{that have head concepts} such as unemployment, inflation, recession and many others that could generate confusion in the annotator regarding the ultimate source of the sentiment.
Some examples of phrases are reported in the online Appendix. 
We ask annotators to assign a numeric value, or score, expressing their subjective sentiment for a specific term. 
The score is a numerical value ranging from $-1$ (very negative) to $1$ (very positive), with a precision of one decimal point.
We make clear to annotators that scoring a term in the interval [$-1$, 0) should be interpreted as a negative outcome for the economic entity in the phrase, and the opposite for scores in $(0,1]$. Values of the score closer to $-1$ and $1$ convey an increasingly negative and positive sentiment, respectively, towards the economic entity. 
The score assigned to each term in the EL represents the median of the distribution of the ten annotators' scores. 


\subsection{Disambiguation}


Eliciting the sentiment of a word is not an easy task because its meaning can be context-specific, in addition to the fact that annotators might interpret differently its economic content. This is evident in our annotations since we found several terms with considerable heterogeneity of the sentiment score across annotators. This ambiguity has the potential to introduce noise in the word score and make the classification in sentiment categories more uncertain. We thus decided to eliminate such ambiguous terms from the EL as explained below. 

From all the terms available in our dictionary, 
we focused on words that showed the largest disagreement, across the 10 annotators, on the sign of the sentiment assigned. The review of this subset of terms is completed by the five authors of this paper who were asked to flag terms that could be considered ambiguous in the context of an economic document. We then ranked the terms and decided to eliminate those that were flagged by at least one of the five annotators as potentially ambiguous. 
After removing 713 ambiguous terms, the EL reduces to 6,670 terms, of which 5,572 have non-zero sentiment, and 3,326 have a negative sentiment. Of this list of terms, 1,382 words are included in the LMD dictionary proposed by \cite{Loughran_McDonald_2011}, most of them carrying negative sentiment. As we discuss more extensively in the following Section, our dictionary expands significantly the set of words relative to LMD, in particular for positive terms such as \textit{stimulate, expansion, recovery}, but also negative terms, such as \textit{slump} or {\it downfall}.
On the other hand, some terms in LMD are not included in our lexicon, mostly accounting terms that are not often used in text related to economic concepts. Some examples are \textit{chargeoffs} and {\it unreimbursed} among others. 
We also compare the EL to the REN lexicon proposed by \cite{renault2017intraday}\footnote{In particular, we employ the $L_1$ lexicon proposed by \cite{renault2017intraday}.} and the SSW dictionary discussed in \cite{shapiro2020measuring}. 
While REN includes 8,000 terms, SSW is a very extensive lexicon of over 18,000 words. In both cases, the sentiment score is model-based which is in contrast to the human annotated scores of the EL and LMD. In the following Section, we investigate in more detail the differences between these lexicons and evaluate empirically their performance in several economic applications of interest.

\section{The battle of lexicons\label{application}}

Table~\ref{comparison_lexicon} shows the number of words in the lexicons that are negative, neutral, and positive, together with the total number of words. In the case of the lexicons that have a numerical score, we convert it into negative/positive categories and consider a score of 0 as neutral. There are several differences between these dictionaries. In terms of coverage, LMD has the smallest set of terms with only 2,709 words, while SSW is the largest with 18,307, although over 10,000 have neutral sentiment. An intermediate coverage is provided by EL and REN which have 6,670 and 8,000 words, respectively. The balance between negative and positive terms is also quite different across lexicons. While LMD includes predominantly negative words, REN is equally split between the two categories, while the negative words for EL and SSW are 60\% and 63\%, respectively, of the total number of positive and negative terms. Hence, EL, REN, and SSW increase the set of words covered in their lexicons and also expand significantly the positive terms relative to LMD, although the negative terms are still predominant. 

\begin{table}[t]
\begin{center}
\begin{tabular}{@{\extracolsep{15pt}}lcccc} 
\\[-1.8ex]\hline 
\hline \\[-1.8ex] 
Dictionary & Negative & Neutral & Positive & Total\\
\hline \\[-1.8ex]
EL & 3326 & 1098 & 2246 & 6670\\
LMD & 2355 & & 354 & 2709\\
REN & 4000 & & 4000 & 8000\\
SSW & 4859 & 10638 & 2810 & 18307\\
\\[-1.8ex]
\hline\hline
\end{tabular}
\caption{Number of negative, neutral, positive and total words in the EL, LMD, REN and SSW dictionaries. }
\label{comparison_lexicon}
\end{center}
\end{table}

\begin{figure}[tbh]
 \centering
 \includegraphics[scale=0.45]{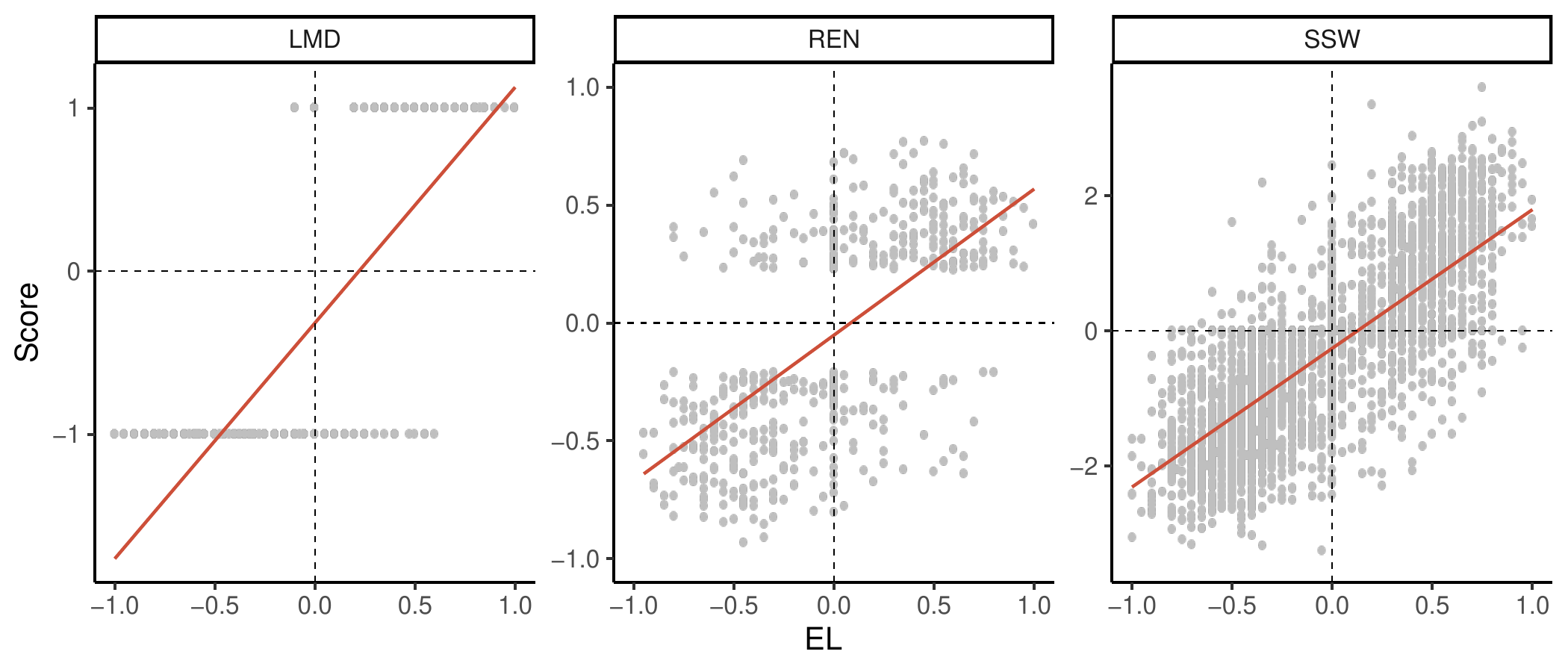}
 \caption{Scatter plots of the sentiment scores of EL against LMD, REN and SSW dictionaries at the word level. We coded the negative/positive categories in LMD as -1 and 1. The points in the scatter plot represent only words that have a score in both lexicons. 
 The red line represents the linear regression fit.}
 \label{fig:scatter1}
\end{figure}

Figure~\ref{fig:scatter1} shows the scatter plot of the sentiment score for EL against the other three lexicons considered, namely LMD, REN, and SSW, for the words that are included in both lexicons. We represent the categories in LMD by a -1 for negative sentiment and 1 for positive. In all cases it is clear that there is positive correlation between the scores across lexicons. In particular, the words categorized as positive in LMD have, to a large extent, also positive sentiment in EL. However, for negative words in LMD we notice that EL assigns positive scores to certain words (more precisely, the lexicons agree on 1080 negative words and disagree on 94). This might indicate that a negative word in the financial and accounting context of LMD might actually have a positive meaning in an economic context. 
When comparing EL and REN, the 2nd and 4th quadrants show that there are several words in which the two lexicons disagree on their sentiments, however the majority of words are in the 1st and 3rd quadrants where the lexicons agree on the sign of sentiment. Although the two dictionaries have 6,670 and 8,000 words, respectively, they have in common only \textcolor{black}{514} terms of which they disagree on the sign in 156 cases\footnote{In the online Appendix, we report the agreement and disagreement between the four competing lexicons as number of words.}.
Instead, the EL and SSW share \textcolor{black}{3,869} words which is a significant overlap relative to the other lexicons. The scatter plot shows that there is high correlation between the scores in the two dictionaries although, as expected, there is also disagreement in the sentiment of many words. More specifically, the dictionaries agree on the sentiment of \textcolor{black}{2,596} (of which 1,381 are negative, \textcolor{black}{762} positive, and \textcolor{black}{453} neutral). 
The disagreement is thus limited to \textcolor{black}{1,273} words. 
Overall it seems that there is agreement between the SSW and EL, at least in the set of words that are in common. This indicates that the procedure used by \cite{shapiro2020measuring}
to define the word sentiment replicates closely the sentiment provided by the human annotators of the EL, at least in its sign. 
However, there is another dimension to consider in the comparison among lexicons. There are many terms that might be included in one lexicon but not in the other given the different ways in which these dictionaries are constructed. 
Continuing with the EL and SSW comparison, there are 2,801 terms that are included in the EL but not in the SSW and these words are mostly negative (60\%) and positive (27\%). Instead, there are \textcolor{black}{14,438} SSW terms that are not included in the EL with 65\% of them classified as neutral, 21\% as negative, and 13\% positive.
 
Summarizing, these comparisons show that there is significant overlap among dictionaries, but also notable differences. They are different in terms of coverage, with LMD having the smallest set of terms and SSW the largest. The set of terms included in the dictionaries are not necessarily incremental in the sense that larger lexicons might not include some words from smaller lexicons. Lastly, the sentiment score associated with the words might be constructed in different ways (human vs machine annotation) which implies different sign and/or magnitude. 

\bigskip

\begin{figure}[tbh]
 \centering
 \includegraphics[scale=0.45]{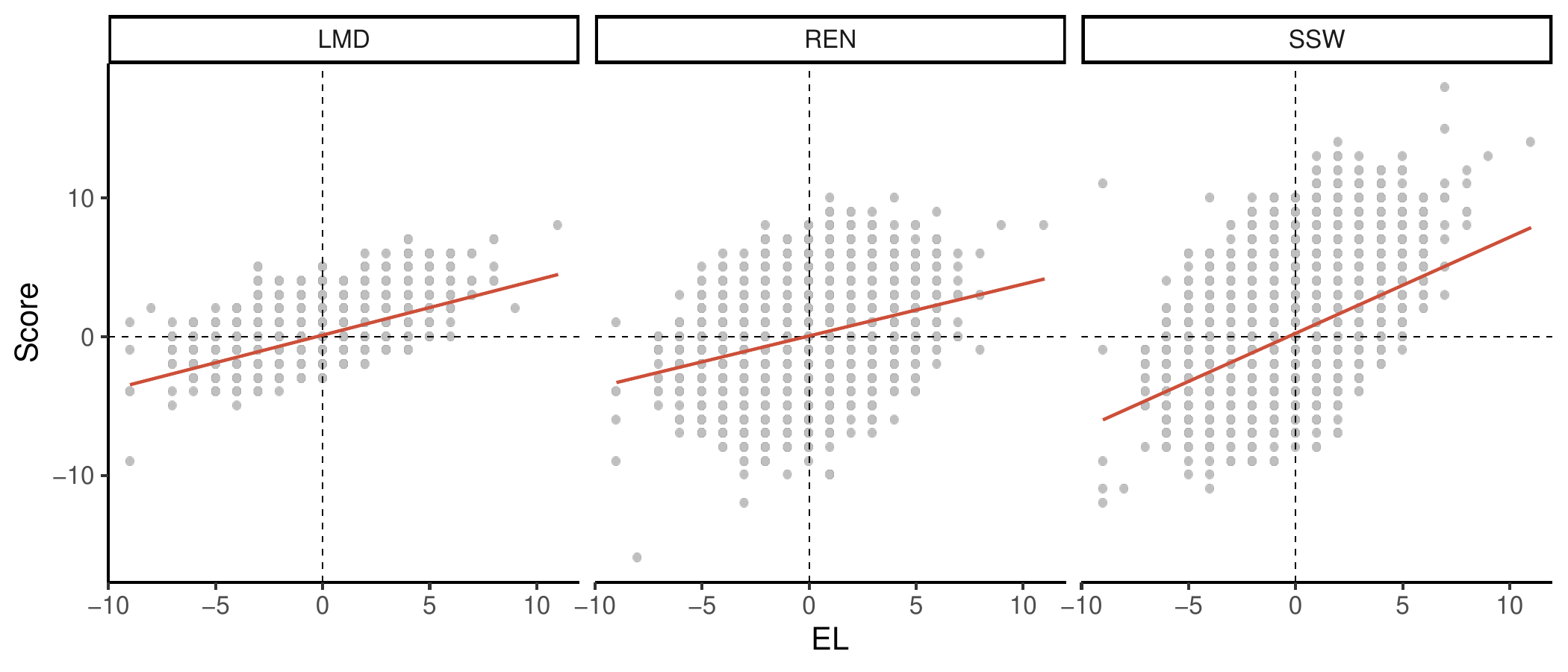}
 \caption{Scatter plots of the sentiment scores of EL against LMD, REN and SSW dictionaries at the sentence level. 
 Individual sentiment scores are coded as negative or positive depending on their sign.
 The red line represents the linear regression fit.}
 \label{fig:scatter2}
\end{figure}

The previous discussion focuses on comparing the lexicons at the word-level. Another way to compare the lexicons is based on the score that they assign to the same sentence discussing an economic concept. Unfortunately, in this case, we do not have a humanly-annotated score for the whole sentence to benchmark the performance of the lexicons as we did for the terms\footnote{An exercise along these lines is performed by \cite{shapiro2020measuring}. They ask 15 annotators to score 800 newspaper articles and then compare the accuracy of lexicons in matching the average human score.}. However, we believe it is still interesting to assess the agreement among dictionaries regarding the sentiment of these sentences and to what extent they disagree. 
We perform two exercises. 
In the first one, we calculate the sentiment score of each sentence as the difference between the sum of positive and negative words. For EL, REN, and SSW we categorize the terms in positive/negative based on the sign of the score. Hence, the sentence has a positive score when it contains more positive, relative to negative, terms according to the lexicon. The different lexicons can thus provide a different score based on the factors discussed earlier, such as its coverage of terms and sentiment classification. Figure~\ref{fig:scatter2} shows the scores for a set of over 3 million sentences\footnote{The set of 3 million sentences contains all the noun-phrases from the newspapers presented in Section \ref{data} that contain at least one of the 355 entities presented in Section 
\ref{sec:econ_concepts}.
} for the EL against the three alternative dictionaries. The sentiments for the four dictionaries are, on average, positively correlated, although there is significant disagreement. In many instances the sign of EL and the other lexicons are opposite, in particular for sentences with few words carrying sentiment. In addition, the slope of the regression line for REN and SSW indicates that, on average, these lexicons tend to provide less extreme scores relative to EL. 

\begin{figure}[tbh]
 \centering
 \includegraphics[scale=0.45]{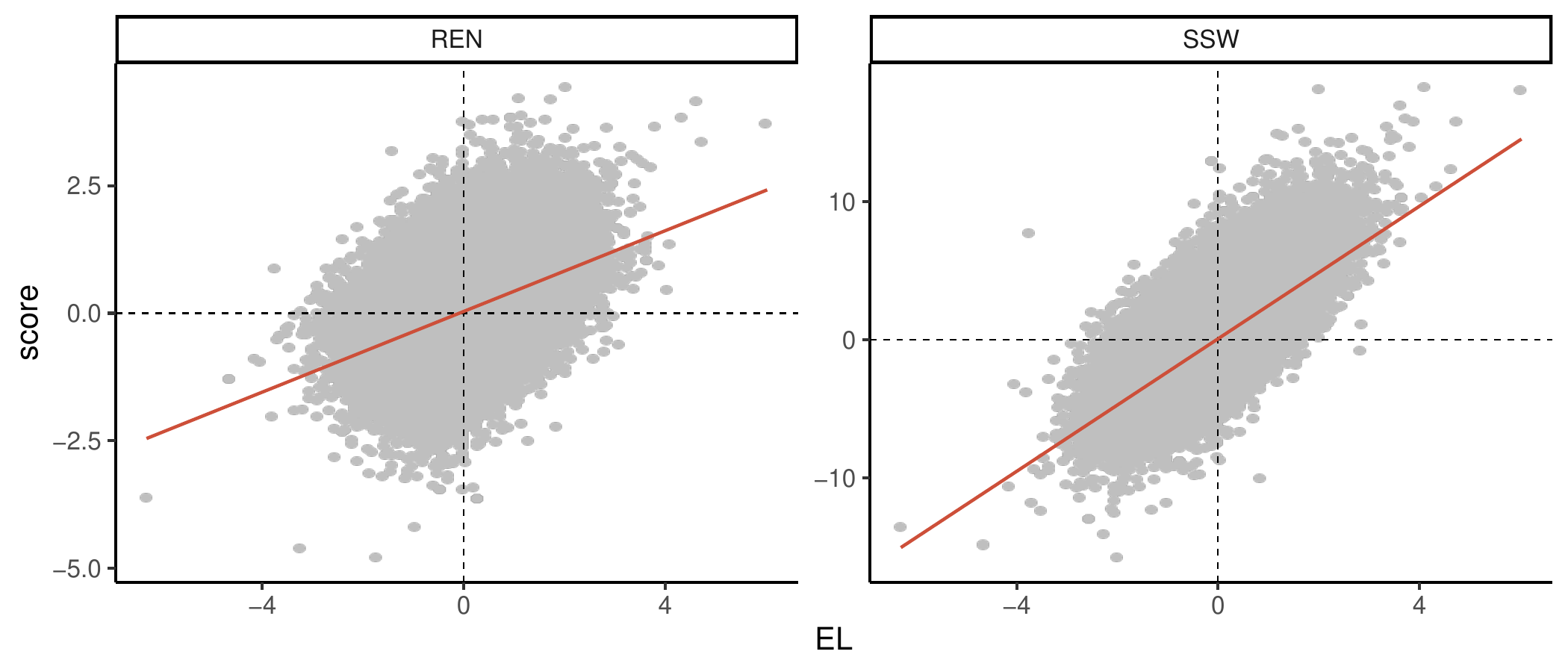}
 \caption{Scatter plots of the sentiment scores of EL against REN and SSW dictionaries at the sentence level. 
 The red line represents the linear regression fit.}
 \label{fig:scatter3}
\end{figure}

The second exercise that we perform uses the continuous sentiment score of EL, REN, and SSW to compute the sentence-level sentiment as the sum of the individual term scores. The picture that emerges in Figure~\ref{fig:scatter3} confirms the positive correlation of the sentiment measures found in Figure~\ref{fig:scatter2}, but also detects considerable disagreement across sentences.
The steeper (relative to the diagonal) regression line suggests that, when considering the value of the sentiment score, SSW delivers more extreme values for the sentence relative to EL. For REN we find the opposite.

\begin{figure}[h]
 \centering
 \includegraphics[scale=0.45]{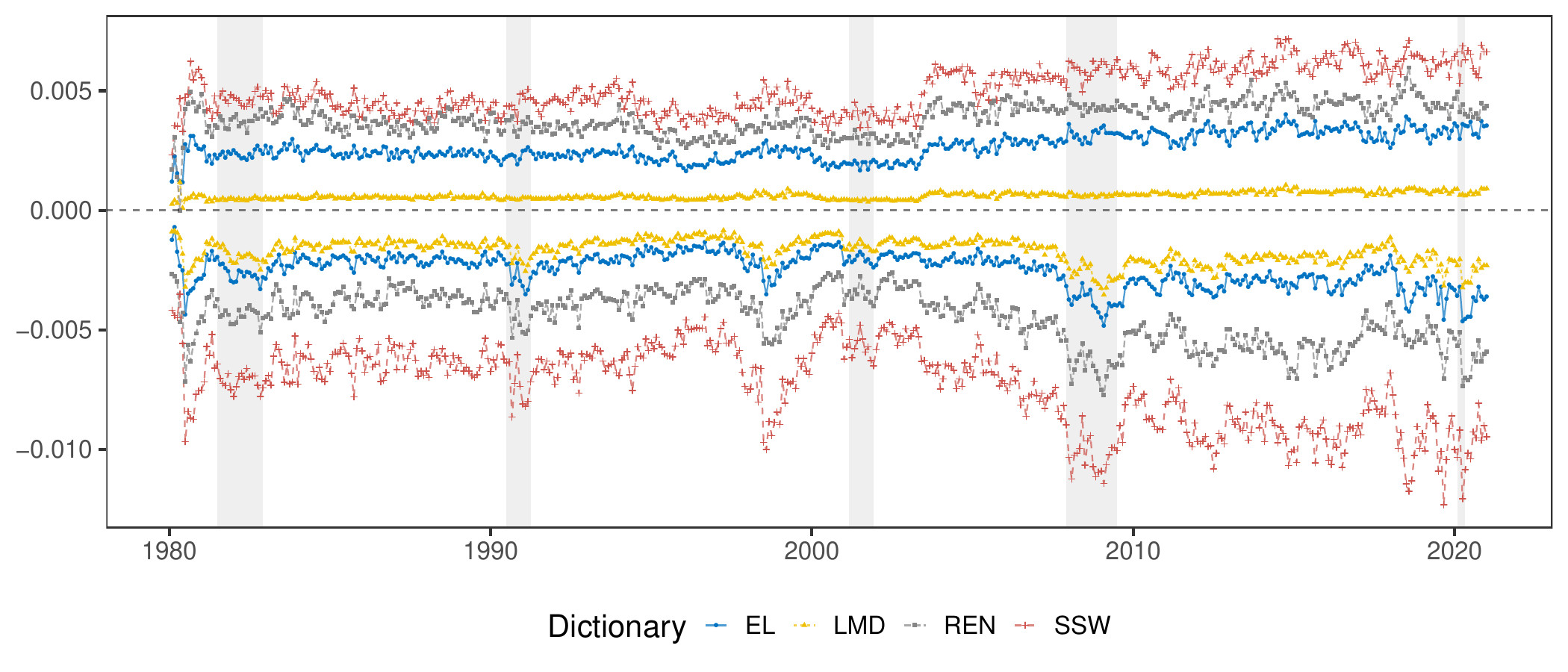}
 \caption{Time series of the number of positive and negative words for the EL, LMD, REN, and SSW dictionaries as a fraction of the total words included in a certain month. The fraction of negative terms is multiplied by -1. }
 \label{fig:timeseries1}
\end{figure}

Another way to investigate the different characteristics of the lexicons is to analyze their behavior over time. Figure~\ref{fig:timeseries1} shows the monthly frequency of positive and negative terms in the four lexicons based on sentences published in newspaper articles that contain an economic term. The results confirm the characteristic of the LMD to have a relatively lower number of negative words with respect to positive ones. The EL increases the frequency of words selected as conveying sentiment, in particular for words with positive sentiment. Relative to EL, both REN and SSW select more positive and negative words in the text, although it seems that the largest difference arises for negative terms.

\section{Economic applications\label{econ_applications}}

In this Section, we discuss some economic applications of sentiment analysis to evaluate the performance of the lexicons presented earlier. In particular, {\color{black} we use a simple measure of sentiment, {\it Economic Pessimism} (EP)\footnote{\color{black} An early application of pessimism to predict stock returns is \cite{Tetlock_2007}. \cite{algaba2020econometrics} is a recent survey of the numerous applications of sentiment analysis in economics and finance.},  that is defined as follows}
\begin{equation}
 EP_{l,t} = - \frac{ \sum_{i=1}^{N_t}  S_{i,l} f_{i,t}  }{N_t},
 \label{eq:EP}
\end{equation}
where $EP_{l,t}$ represents the EP using lexicon $l$ in month $t$, $N_t$ is the total number of words observed in month $t$, 
{\color{black} $S_{i,l}$ denotes the sentiment score  of word $i$ in lexicon $l$, and $f_{i,t}$ represents the frequency count of word $i$ in month $t$. The sentiment score $S_{i,l}$ can be equal to $\pm$1 for lexicons that classify terms in positive and negative categories, or a value in the interval [-1,1] for fine-grained dictionaries\footnote{\color{black} To make results comparable for the LMD and the other lexicons we converted all fine-grained scores to either 1 (positive) or -1 (negative). Hence, the term $S_{i,l}$ takes value $\pm$1 for all lexicons $l$ included in the analysis. Our findings are not sensitive to this choice and are similar when using the fine-grained scores. For example, the $EP$ measure using the fine-grained EL scores rather than the derived positive/negative categories is 0.97. However, the fine-grained nature of these lexicons might be more relevant in the context of more sophisticated sentiment measures as the one proposed in \cite{barbaglia2020forecasting} among others. }.} 
For lexicons that divide terms into positive and negative groups, $EP_{l,t}$ represents the difference between the total frequency of negative and positive words in a certain month (as seen in Figure~\ref{fig:timeseries1}), with positive values of the measure indicating the predominance of negative terms in economic news. 
We calculate the $EP_{l,t}$ measure on all economic-relevant phrases extracted from news published in the six main US newspapers mentioned earlier from January 1st, \textcolor{black}{1980} to December 31st, \textcolor{black}{2020}.
To facilitate the comparison across lexicons, we standardize the economic pessimism measures to have mean zero and variance equal to one. 
The goal of this Section is to  {\color{black} evaluate the performance of the different lexicons in measuring sentiment based on a simple and transparent pessimism measure. There are  alternative and more sophisticated dictionary-based approaches to measure sentiment that we could have used, such as the VADER approach proposed by \cite{hutto2014vader} and the aspect-based methodology of \cite{barbaglia2020forecasting}. These measures are likely to provide more accurate sentiment measures, but their nonlinear nature makes it difficult to disentangle the sources of the differential performance of lexicons relative to the linear EP indicator.  
}

\begin{figure}[]
 \centering
 \includegraphics[scale=0.48]{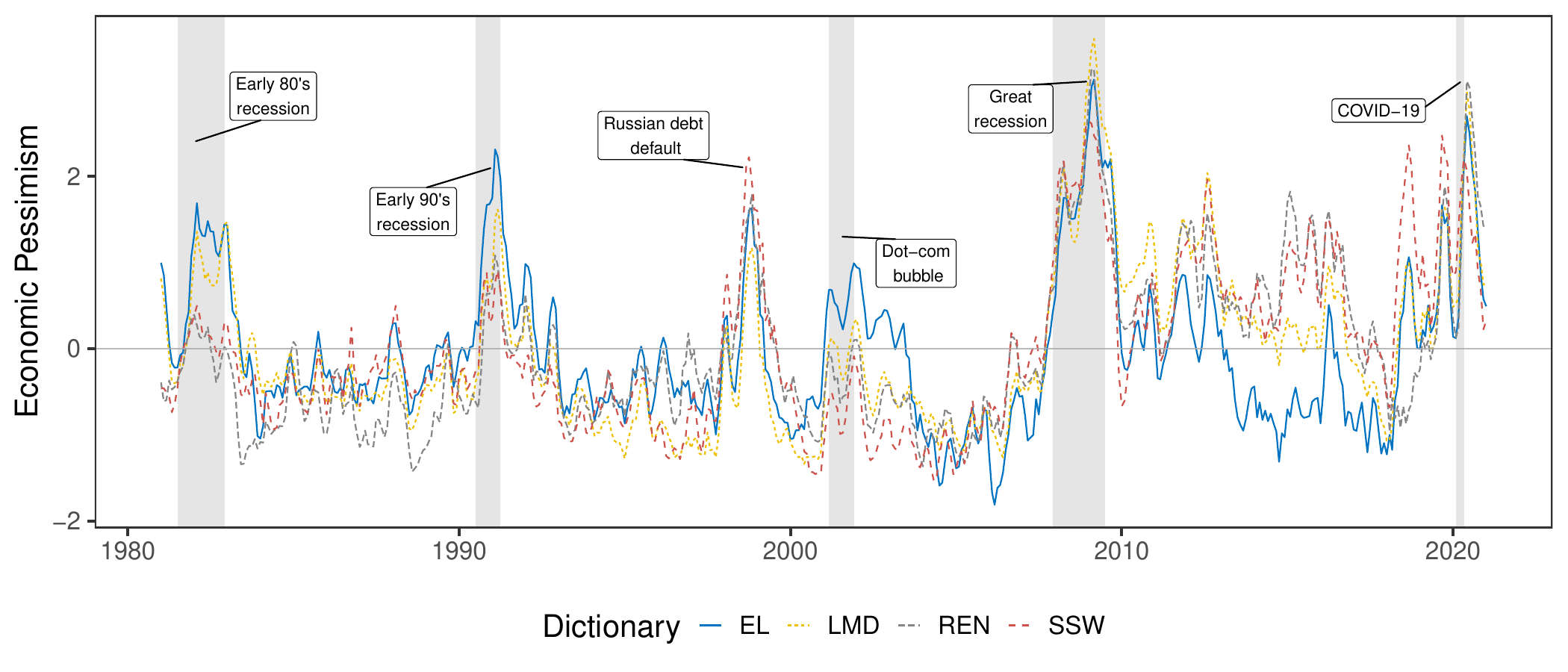}
 \caption{Time series of \textit{Economic Pessimism} by lexicon (i.e., $EP_{EL}$, $EP_{LMD}$, $EP_{SW}$ and $EP_{REN}$) at the monthly frequency. 
 \textcolor{black}{
The series are standardized to have mean zero and variance one, and smoothed with a three-month rolling moving average to appreciate the differences among indicators. 
 The grey areas denote the recessions dated by the NBER.
  } }
 \label{fig:sentvar}
\end{figure}

Figure \ref{fig:sentvar} displays the monthly time series of the pessimism measure using the EL, LMD, REN and SSW lexicons. 
We obtain the monthly time series by averaging sentence-level scores within the month, and standardize the series to have mean zero and variance equal to one. 
Overall, there is significant co-movement in the four series and they seem to capture well
the business cycle fluctuations. The largest correlation of $EP_{EL}$ is relative to $EP_{LMD}$ and equals 
0.82\footnote{Notice that the high correlation is accentuated by the monthly aggregation. Considering the sentiments at the daily frequency the correlation between $EP_{EL}$ and $EP_{LMD}$ would be 0.66.
}, while the correlations with $EP_{SSW}$ and $EP_{REN}$ are 
0.64 and 
0.66, respectively. 
$EP_{LMD}$ has also a correlation of 
0.84 with $EP_{SSW}$ and 
0.83 with $EP_{REN}$. 
However, there is also enough independent variation in the measures deriving from the different lexicons, and in particular for EL.

For example, there are sharper peaks for $EP_{EL}$ in the early 1980s and the early 1990s in coincidence with recessionary periods. In addition, during and following the 2001-02 recession $EP_{EL}$ indicates a significant and more persistent pessimism relative to the other measures up to the end of 2004. The pessimism generated by financial events is also captured by the $EP$ measures, such as during the Russian financial crisis of 1998 and the Great Recession of 2008-09. An interesting period in which the sentiment measures diverge is between 2013 and 2016. While the $EP_{EL}$ indicated optimism in economic news, the other measures were notably pessimistic, and significantly so.
Below, we will show that the predictive performance varies largely when the measure of EP is built with the different lexicons\footnote{In the online Appendix, we also consider the lexicon proposed by \cite{correa2021sentiment} to measure sentiment in financial stability reports.}.


\subsection{Pessimism, lexicons, and uncertainty}

In this Section, we evaluate if and to what extent the lexicons determine different comovement of the $EP$ alongside two widely used measures of economic and financial uncertainty. 
The first indicator that we consider is the CBOE Volatility Index (VIX) which is calculated based on option prices and is designed to capture the market expectation of future volatility. The second indicator is the Economic Policy Uncertainty (EPU) Index proposed by \cite{Bloom_Baker_Scott_2016}. As discussed earlier, EPU is a text-based measure that is calculated by counting the number of newspaper articles that contain the words economy/economic, uncertain/uncertainty, and a policy-related term (e.g., Congress, tax).

\begin{figure}[]
 \centering
 \includegraphics[scale=0.45]{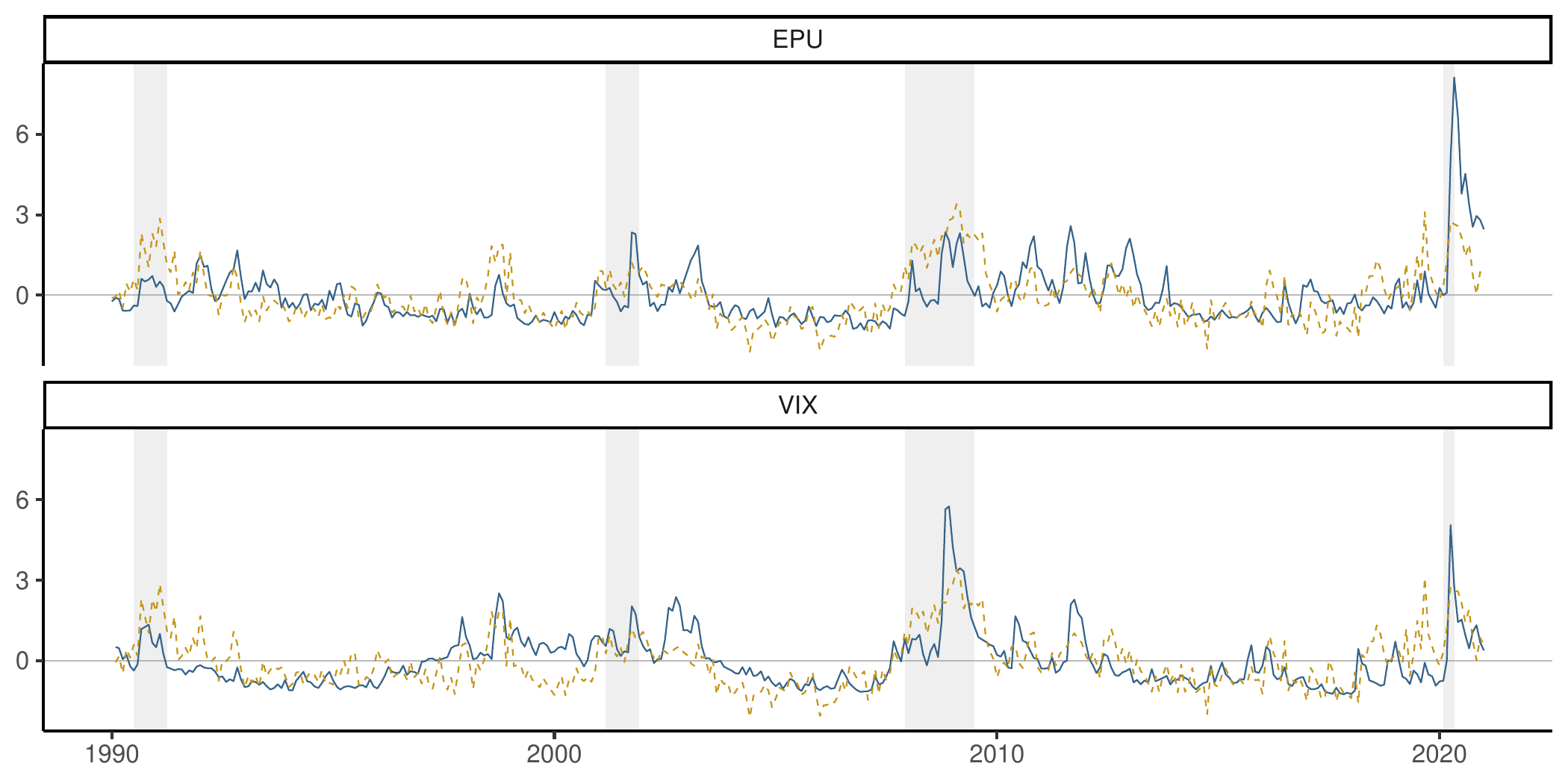}
 \caption{ EPU and VIX (blue solid line in the top and bottom panels, respectively) plotted together with the $EP_{EL}$ sentiment measures (black dashed line). The series are standardized to have mean zero and variance one. The vertical areas represent the NBER recession periods. }
 \label{fig:epu_vix}
\end{figure}

In Figure \ref{fig:epu_vix} we show the monthly time series of the EPU and VIX\footnote{We obtain both indicators at the monthly frequency from the Saint Louis Fed FRED database available at \url{https://fred.stlouisfed.org/} with tickers {\tt VIXCLS} and {\tt USEPUINDXD}.} indicators together with the $EP_{EL}$. Overall, the three series show significant co-movements with higher values during recessions and lower values during expansions, but also similar low-frequency fluctuations. More precisely, $EP_{EL}$ has a correlation of 0.63 and 0.56 with VIX and EPU, respectively\footnote{The correlation of the $EP_{LMD}$ measure with VIX and EPU are 0.54 and 0.56, respectively. The lowest correlations are obtained for $EP_{SSW}$ that are estimated at 0.37 and 0.33, respectively.}. The positive correlation with the VIX is not surprising since the pessimism measure is constructed on a large set of text that includes also financial concepts and makes it responsive to financial market events. 
A number of interesting facts emerge from these findings.
First, the $EP_{EL}$ represents a directional indicator that captures the optimism and pessimism embedded in economic news at a given time. As such, EL does not target explicitly to measure economic and financial uncertainty as EPU and VIX are designed to do. We believe that the high correlation occurs because uncertainty is driven, to a large extent, by negative economic and financial shocks that are likely to increase the pessimism measure.
 Another factor to consider is that EL includes words that express uncertainty and these terms are assigned a negative score by the human annotators. 
 If this is the case, our pessimism indicator might be considered a proxy for uncertainty and thus explain the correlation with VIX and EPU. 
 To evaluate this hypothesis, we match the words in the EL with the 297 words that \cite{Loughran_McDonald_2011} classifies as expressing ``Uncertainty''. However, we found that none of these words are included in our lexicon which seems not to support this hypothesis.


As discussed earlier, the lexicons produce measures of pessimism that are correlated because many terms are in common and also because they classify similarly the sentiment of the terms. On the other hand, the correlation is not extremely large which suggests also significant differences in word coverage and sentiment scores. 
In order to evaluate more formally the ability of EP to explain the uncertainty measures, we estimate the following regression model: 

\begin{equation}
 Y_t = \beta_0 + \sum_{i=1}^p \beta_i Y_{t-i} + \delta EP_{EL,t} + \gamma EP_{l,t} + \epsilon_t,
 \label{eq:eq_uncertainty}
\end{equation}
where $Y_t$ represents either EPU or VIX in month $t$, and $l$ is either LMD, REN, or SSW. The specification thus explains the current value of an uncertainty measure based on its own lags and the EP measures computed based on the EL and an alternative lexicon. The goal of the analysis is thus to evaluate if the LMD, REN, or SSW lexicons provide a measure of pessimism that has additional explanatory power relative to EL. The findings provided in Table~\ref{tab:uncertainty} show that for both EPU and VIX the pessimism measure computed based on the EL lexicon is strongly significant. In addition, it remains significant even when adding the EP measures calculated based on the other dictionaries, which are not significant in any case. The estimate of the effect of $EP_{EL}$ declines slightly when $EP_{LMD}$ is added, but the effect of $EP_{LMD}$ is not significant at 10\%. In this sense, the EL lexicon seems to provide a reliable coverage and sentiment classification that outperforms the other lexicons in this application.


\begin{table}[!htbp] \centering 
\begin{tabular}{@{\extracolsep{-2pt}}lccccccccc} 
\\[-1.8ex]\hline \hline \\[-1.8ex] 
 & \multicolumn{4}{c}{\textit{$EPU_t$}} &\hspace{0.1cm} & \multicolumn{4}{c}{\textit{$VIX_t$}}
\\ \cline{2-5}\cline{7-10}\\[-1.8ex] \\ 
 $EP_{EL,t}$ & 15.692$^{***}$ & 9.509$^{*}$ & 14.751$^{***}$ & 13.630$^{**}$ && 1.455$^{***}$ & 1.388$^{***}$ & 1.339$^{***}$ & 1.154$^{**}$ \\ 
 & (0.240) & (5.521) & (5.460) & (5.665) && (0.541) & (0.412) & (0.479) & (0.470) \\ 
 $EP_{LMD,t}$ & & 7.930 & & & & & 0.081 & & \\ 
 & & (5.025) & & & & & (0.465) & & \\ 
 $EP_{SSW,t}$ & & & 1.301 & & & & & 0.149 & \\ 
 & & & (3.446) & & && & (0.278) & \\ 
 $EP_{REN,t}$ & & & & 3.311 & & & & & 0.482 \\ 
 & & & & (2.977) & & & & & (0.314) \\ 
\hline \\[-1.8ex] 
Obs. & 431 & 431 & 431 & 431 & & 368 & 368 & 368 & 368 \\ 
R$^{2}$ & 0.488 & 0.495 & 0.489 & 0.490 & & 0.758 & 0.758 & 0.758 & 0.760 \\ 
$\overline{R}^{2}$ & 0.485 & 0.490 & 0.484 & 0.486 & & 0.754 & 0.754 & 0.754 & 0.756 \\ 
\hline 
\hline \\[-1.8ex] 
\multicolumn{4}{l}{{\footnotesize \textit{Notes}: $^{***}$ 1\%, $^{**}$ 5\%, $^{*}$ 10\% significance.}} 
\end{tabular}
\caption{Estimation results for the model in Equation~\eqref{eq:eq_uncertainty}. The values in parenthesis are Newey-West corrected standard errors. We include an intercept and lagged values of the dependent variables (2 for EPU and 4 for VIX) that are omitted from the table to save space. } 
\label{tab:uncertainty}
\end{table}

\subsection{Pessimism and consumer sentiment}

\begin{figure}[]
 \centering
 \includegraphics[scale=0.55]{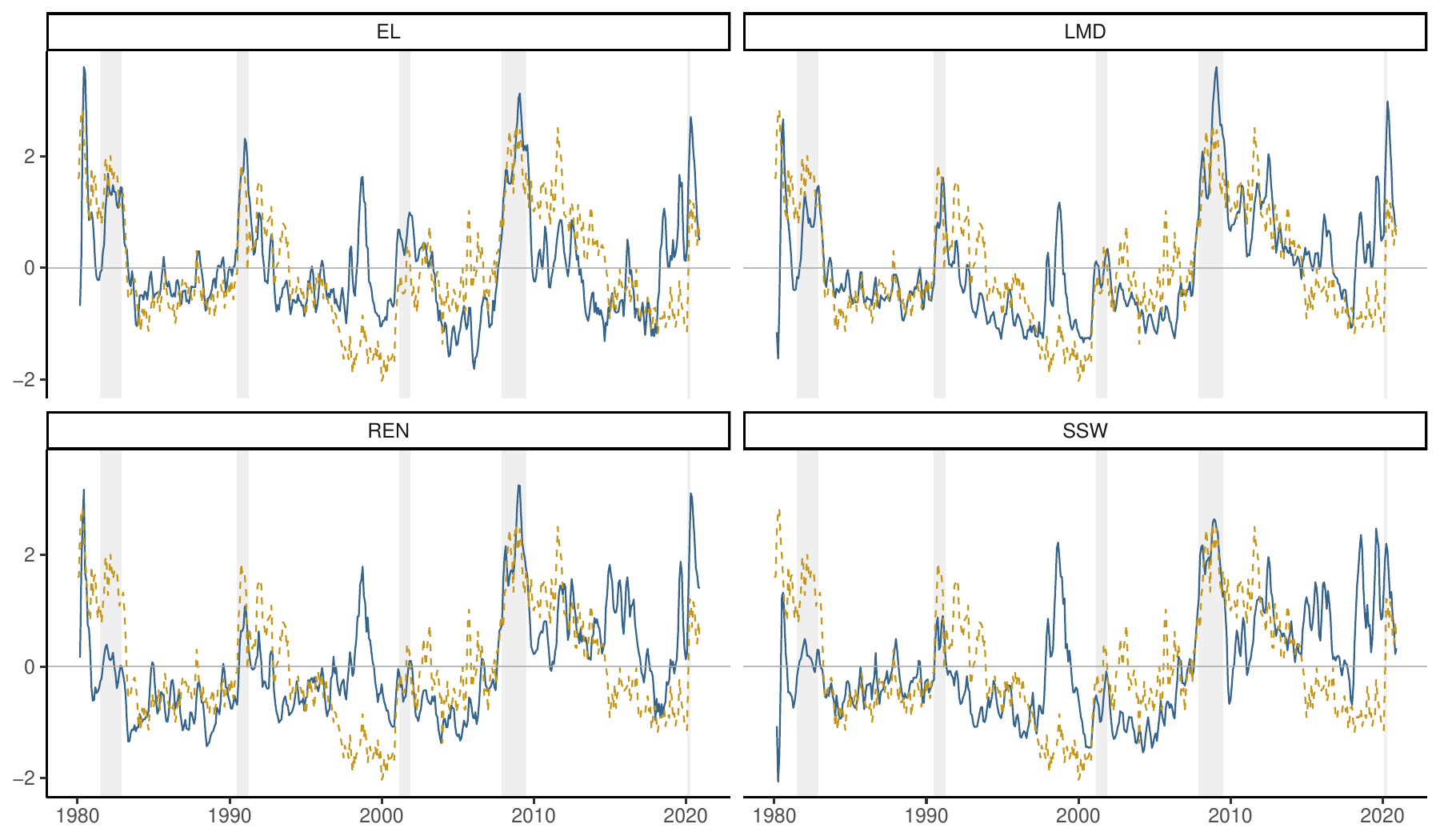}
 \caption{Michigan Consumer Sentiment Index (black dashed line) plotted together with the $EP$ sentiment measures built using the EL, RLM, REN and SSW lexicons (blue solid lines). We changed the sign of the consumer sentiment to match the direction of pessimism.
 All series are standardized to have mean zero and variance one, \textcolor{black}{and smoothed with a three-month rolling moving average to appreciate the differences among indicators.}. The grey areas represent the NBER recession periods. }
 \label{fig:umcsent}
\end{figure}

There is a long history in the economics literature of surveying consumers and businesses regarding their view of the state of the economy \citep[see][]{curtin1982indicators}. The Michigan Consumer Sentiment Index (MCSI) measures the consumers' confidence in the economy and it is considered a leading indicator of the business cycle. There are several papers that evaluate the relationship between the sentiment obtained from textual data and consumer sentiment \citep{ larsen2021news, shapiro2020measuring}.  
In this Section, we compare our EP measure to the MCSI\footnote{We downloaded the series from the Saint Louis Fed FRED database for the ticker {\tt UMCSENT}.} in order to evaluate to what extent there is a connection between the two sentiment measures.
Figure~\ref{fig:umcsent} shows the time series of consumer sentiment together with economic pessimism computed using the 4 lexicons discussed earlier. The EP measures seem to reflect closely the fluctuations in the sentiment of consumers, particularly during recessions. However, there are also periods in which the measures diverge. One example is during the financial turbulence in the summer of 1998 {\color{black} triggered by the Russian devaluation of the ruble and the fallout on the US financial system}. During this period there was a rapid increase of the {\color{black} text-based pessimism driven by financial news discussing the events that, however,}
had hardly any effect on consumer sentiment. This is due to the fact that the EP is constructed based on text that contains both economic and financial terms and thus captures events discussing the financial sector. In terms of correlation, the EP measure that co-varies the most with consumer sentiment is LMD (-0.61), followed by EL (-0.55), REN (-0.47), and SSW (-0.29).


In Table~\ref{tab:umcsent} we provide the estimation results for Equation~\eqref{eq:eq_uncertainty} when the dependent variable is the consumer sentiment (denoted by $MCSI_t$). Our findings suggest very similar results to the case of VIX and EPU, that is, when $EP_{EL}$ is included in the regression the pessimism measures based on the other dictionaries are insignificant. Only when including $EP_{LMD}$ the coefficient of $EP_{EL}$ declines by approximately one standard deviation, but $EP_{LMD}$ is not significant at 10\%. 
A one standard deviation increase in pessimism is associated with a 1.1-1.4 point decline in consumer sentiment.

\begin{table}[!htbp] \centering 
\begin{tabular}{@{\extracolsep{20pt}}lcccc} 
\\[-1.8ex]\hline 
\hline \\[-1.8ex] 
 & \multicolumn{4}{c}{\textit{$MCSI_t$}} \\ \cline{2-5}\\[-1.8ex] 
 $EP_{EL,t}$ & $-$1.414$^{***}$ & $-$1.105$^{***}$ & $-$1.459$^{***}$ & $-$1.303$^{***}$ \\ 
 & (0.240) & (0.295) & (0.264) & (0.249) \\ 
 $EP_{LMD,t}$ & & $-$0.429 & & \\ 
 & & (0.319) & & \\ 
 $EP_{REN,t}$ & & & 0.069 & \\ 
 & & & (0.222) & \\ 
 $EP_{SSW,t}$ & & & & -0.202 \\ 
 & & & & (0.257) \\ 
\hline \\[-1.8ex] 
Obs. & 486 & 486 & 486 & 486 \\ 
R$^{2}$ & 0.908 & 0.908 & 0.908 & 0.908 \\ 
$\overline{R}^{2}$ & 0.906 & 0.907 & 0.906 & 0.906 \\ 
\hline 
\hline \\[-1.8ex] 
\multicolumn{4}{l}{{\footnotesize \textit{Notes}: $^{***}$ 1\%, $^{**}$ 5\%, $^{*}$ 10\% significance.}} 
\end{tabular} 
\caption{Estimation results for the model in Equation~\eqref{eq:eq_uncertainty} and dependent variable the University of Michigan consumer sentiment, $MCSI_t$. The values in parenthesis are Newey-West corrected standard errors. We include an intercept and six lagged values of the dependent variable that are omitted from the table to save space. }
\label{tab:umcsent}
\end{table}

\subsection{Forecasting economic recessions}

The previous findings show that the pessimism measure is closely related to measures of uncertainty and consumer sentiment. Another relevant application is to use the pessimism indicator to predict real macroeconomic outcomes and the business cycle. In particular, we consider the ability of the EP measure to predict a binary variable, denoted $Recession_t$, that takes value 1 if the economy is in a recession in month $t$ and 0 during expansions as declared by the NBER Business Cycle Dating Committee. In addition to the text-based sentiment, we control for the state of the economy by including the slope of the yield curve, defined as the difference between the 10-year and the 3-month Treasury yield (denoted by $Spread_t$), and a proxy for the business cycle proposed by \cite{aruoba2009real} (denoted $ADS_t$). The ADS indicator is the outcome of a dynamic factor model based on a small set of macroeconomic variables whose frequencies range from weekly to quarterly. Hence our forecasting model is:
\begin{equation}
 P\left(Recession_{t+h} = 1\right) = F\left(\beta_0 + \beta_1 Spread_t + \beta_2 ADS_t + \gamma EP_{l,t}\right),
 \label{eq:logit1}
\end{equation}
where $h$ indicates the forecast horizon, $l$ denotes the dictionary used to compute EP, and $F(\cdot)$ is the logistic distribution. The model is estimated by maximum likelihood, using monthly data from January 1980 to December 2020.
Table \ref{tab:logit1} reports the regression results for the case $h=3$. 
The findings show that the estimated coefficients for the EP measures calculated with the different dictionaries are all statistically significant at 1\% thus indicating that they provide valuable incremental information relative to the term spread and the ADS index. The model based on the $EP_{EL,t}$ achieves best performance, in terms of log-likelihood and AIC, relative to the cases when EP is constructed using the LMD, REN, and SSW lexicons. However, the fact that all EP measures are significant leaves open the question of whether these measures are complementary in forecasting recessions relative to the case that one of them leads all others in terms of forecasting power. 

\begin{table}[H] \centering 
\begin{tabular}{@{\extracolsep{20pt}}lcccc} 
\\[-1.8ex]\hline \hline 
\\[-1.8ex] & \multicolumn{4}{c}{$RECESSION_{t+3}$} \\ \cline{2-5} \\[-1.8ex] 
 $EP_{EL,t}$ & 0.922$^{***}$ & & & \\ 
 & (0.156) & & & \\ 
 $EP_{SSW,t}$ & & 0.378$^{***}$ & & \\ 
 & & (0.140) & & \\ 
 $EP_{LMD,t}$ & & & 0.710$^{***}$ & \\ 
 & & & (0.152) & \\ 
 $EP_{REN,t}$ & & & & 0.424$^{***}$ \\ 
 & & & & (0.156) \\ 
\hline \\[-1.8ex] 
Obs. & 491 & 491 & 491 & 491 \\ 
Log Lik. & $-$134.830 & $-$148.836 & $-$141.400 & $-$148.888 \\ 
AIC & 277.659 & 305.672 & 290.799 & 305.776 \\ 
\hline 
\hline \\[-1.8ex] 
\multicolumn{4}{l}{{\footnotesize \textit{Notes}: $^{***}$ 1\%, $^{**}$ 5\%, $^{*}$ 10\% significance.}} 
\end{tabular} 
 \caption{Estimation results for the Logistic regression in Equation~\eqref{eq:logit1} for the case $h=3$ and $l \in \{EL, LMD, REN, SSW\}$. Results for $Spread_t$ and $ADS_t$ are omitted to save space. Standard errors are reported in parenthesis.} 
 \label{tab:logit1} 
\end{table}

To answer this question we estimate the following model: 
\begin{equation}
 P\left(Recession_{t+h} = 1\right) = F\left(\beta_0 + \beta_1 Spread_t + \beta_2 ADS_t + \gamma_{EL} EP_{EL,t} + \gamma_{l} EP_{l,t} \right),
 \label{eq:logit2}
\end{equation}
where we include both the EP measure calculated using EL and one of the remaining three lexicons, with $l \in \{LMD, REN, SSW\}$. The results in Table~\ref{tab:logit2} show that when the $EP_{EL}$ is included jointly with any of the other three measures, it absorbs the significance and makes the alternative measures irrelevant. This result confirms the findings in the earlier sections: EL seems to produce measures of pessimism that perform better, relative to alternative lexicons, in several applications of interest to economists. 
A question not addressed in this paper is to what extent the superiority of the EL extends to other measures of sentiment or different types of sentiment analysis.

\begin{table}[H] \centering 
\begin{tabular}{@{\extracolsep{25pt}}lccc} 
\\[-1.8ex]\hline \hline 
\\[-1.8ex] & \multicolumn{3}{c}{$RECESSION_{t+3}$} \\ \cline{2-4} \\[-1.8ex] 
 $EP_{EL,t}$ & 1.021$^{***}$ & 0.889$^{***}$ & 1.141$^{***}$ \\ 
 & (0.202) & (0.259) & (0.225) \\  
 $EP_{SSW,t}$ & $-$0.149 & & \\ 
 & (0.185) & & \\ 
 $EP_{LMD,t}$ & & 0.038 & \\ 
 & & (0.243) & \\ 
 $EP_{REN,t}$ & & & $-$0.294 \\ 
 & & & (0.206) \\ 
 & & & \\ \hline \\[-1.8ex] 
Obs. & 491 & 491 & 491 \\ 
Log Lik. & $-$134.494 & $-$134.818 & $-$133.791 \\ 
AIC & 278.988 & 279.635 & 277.582 \\ 
\hline 
\hline \\[-1.8ex] 
\multicolumn{4}{l}{{\footnotesize \textit{Notes}: $^{***}$ 1\%, $^{**}$ 5\%, $^{*}$ 10\% significance.}} 
\end{tabular} 
 \caption{Estimation results for the Logistic regression in Equation~\eqref{eq:logit2} for the case $h=3$ and $l \in \{LMD, REN, SSW\}$. Results for $Spread_t$ and $ADS_t$ are omitted to save space. Standard errors are reported in parenthesis.} 
 \label{tab:logit2} 
\end{table}

\begin{figure}[h!]
 \centering
 \includegraphics[scale=0.45]{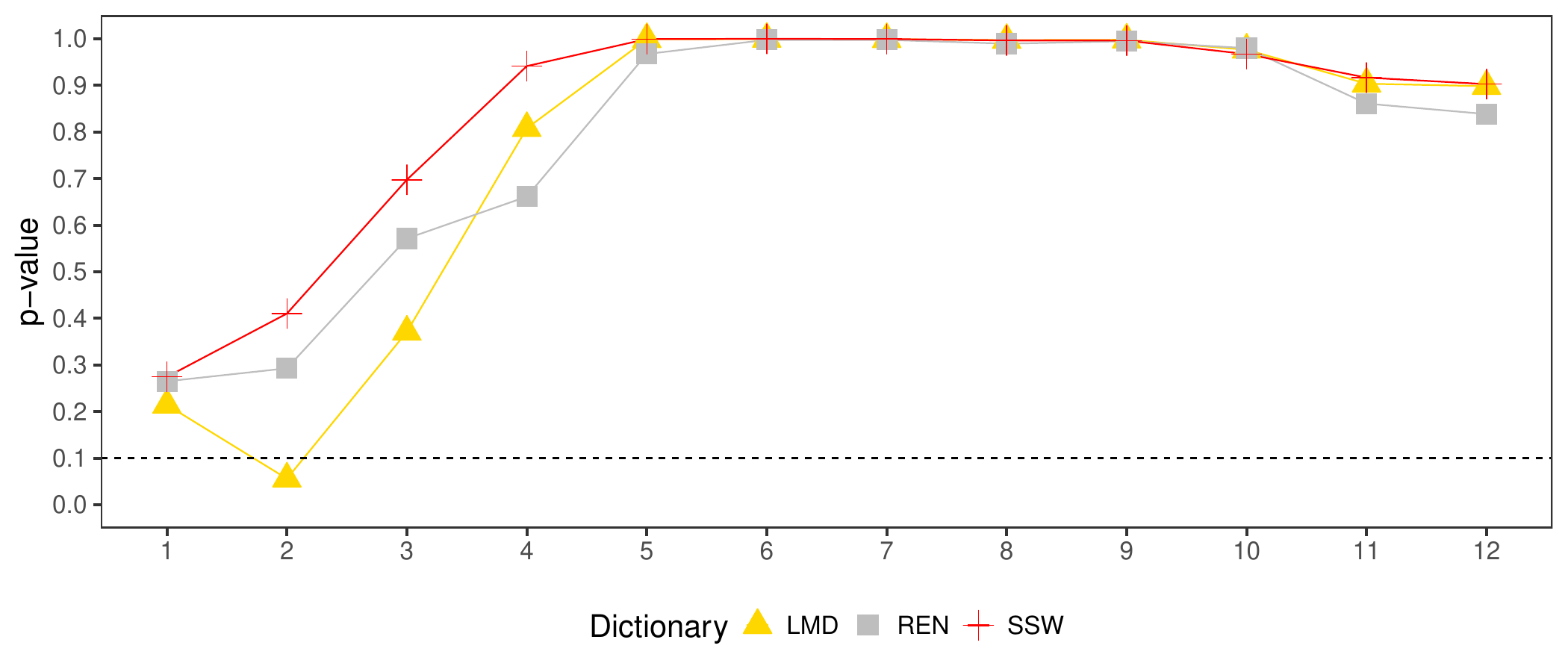}
 \\
\caption{One-sided p-value for the null hypothesis that the Area Under the Curve (AUC) for model (3) using the EL dictionary is larger or equal to the alternative dictionary, against the alternative that it is smaller. Horizon $h$ goes from 1 to 12 months. }
 \label{fig:AUC_h}
\end{figure}

The previous discussion focuses on one specific forecast horizon. To evaluate the performance of the lexicons more generally, we calculate the Area Under the Curve (AUC) for the regression model in Equation~\eqref{eq:logit1} for the forecast horizon $h$ ranging from 1 to 12. 
{\color{black} Figure~\ref{fig:AUC_h} shows the p-value for the null hypothesis that the AUC of the model based on EL is greater or equal to the AUC for the economic pessimism based on the alternative dictionaries. 
The evidence suggests that at most horizons we do not reject the null hypothesis, except for LMD at $h=2$. This indicates that at most horizons the economic pessimism constructed on the EL lexicon provides better performance, at least in terms of larger AUC, in forecasting the occurrence of a recession at $t+h$. } 

\subsection{Explaining the differences}

\color{black}

The goal of this Section is to understand the driving factors of the superior performance of the EL relative to the other dictionaries. As we have discussed earlier, the EP measure might differ across lexicons because of differences in the word coverage as well as due to heterogeneity in the sentiment scores of the words. To isolate these two effects, we create EP measures for modified versions of lexicon $l$ that use information from the EL dictionaries.  More specifically, the lexicon $l$ (with $l$ being LMD, REN or SSW) and EL  share a set of terms that they both consider either positive or negative, but they also disagree  on the sentiment of certain words. In addition, the EL includes terms that are not part of the alternative lexicon due to differences in coverage. We thus construct the following modified sentiment measures for lexicon $l$:
\begin{itemize}
    \item $EP^{disagree}_{l,t}$: 
    EP calculated on a modified lexicon $l$ where the sentiment of terms in common with $EL$ but with opposite scores are assigned  the sentiment in $EL$. All other terms maintain the score of lexicon $l$;
    \item $EP^{onlyEL}_{l,t}$: 
    EP calculated on a modified lexicon $l$ in which terms that are neutral or missing in lexicon $l$, but are part of $EL$, are included in the modified lexicon $l$ with the $EL$ sentiment score.
\end{itemize}
These EP measures are highly correlated with the EP calculated on the original lexicon since only a relatively small fraction of terms change sentiment scores or are included in the new measures. To address this issue, we include these variables in difference relative to the EP calculated on the original lexicon, that is,
we define $\Delta EP^{disagree}_{l,t} =  EP_{l,t}^{disagree} - EP_{l,t}$ and  $\Delta EP^{onlyEL}_{l,t} = EP_{l,t}^{onlyEL} - EP_{l,t}$. We then consider a model for the probability of a recession happening at month $t+h$ that includes the EP calculated with lexicon $l$ and the two variables introduced above, that is:
\begin{align}
 \label{eq:eq_log}
 P\left(Recession_{t+h} = 1\right) &= F\bigl(\beta_0 + \beta_1 Spread_t + \beta_2 ADS_t +\\
 & \mbox{\hspace{0.5cm}} \gamma_{l} EP_{l,t} + \gamma_{d} \Delta EP^{disagree}_{l,t} + \gamma_{o} \Delta EP^{onlyEL}_{l,t} \bigr).\notag
\end{align}
The goal of this specification is to evaluate if nesting information from the EL lexicon in the other lexicons improves the explanatory power of the model relative to simply considering the pessimism based on the lexicon $l$. In particular, we expect that the $ \Delta EP^{disagree}_{l,t}$  will be significant if EL provides more accurate sentiment scores for the terms in disagreement. A positive value of $\Delta EP^{disagree}_{l,t}$ indicates that the pessimism based on the modified lexicon is higher relative to using the original lexicon. The only reason for such a difference to occur is the change of score of the terms in disagreement with EL. In this case, a positive coefficient $\gamma_d$ would predict a higher probability of a recession at $t+h$.
On the other hand, we expect $\Delta EP^{onlyEL}_{l,t}$ to be significant if the difference in word coverage between EL and dictionary $l$ provides forecasting power. A positive value of $\Delta EP^{onlyEL}_{l,t}$ means that the terms added to the lexicon $l$ (that are included in EL) increase the pessimism relative to the EP measure calculated on lexicon $l$ that does not include them. Also in this case, a positive coefficient $\gamma_{o}$ can be interpreted as an increase in the probability of a recession at $t+h$ due to these additional words at time $t$. \\
Table~\ref{tab:tab1} shows that the disagreement term is significant for LMD and SSW.  We interpret this fact as evidence that the predictive content of $EP_{l,t}$ can be increased by switching the sentiment score of a subset of terms to the values they are assigned in the EL lexicon. Expanding the coverage of the lexicon $l$ with words that are only included in EL is beneficial for all lexicons considered and it has particularly significant effects for LMD and REN. Coefficients associated with $EP_{l,t}$ are significant in all cases since this measure of pessimism includes the terms in agreement with EL as well as the terms that are not included in EL\footnote{In a separate exercise, we also control for $EL$ by including $EP_{EL,t}$ as an additional regressor in (\ref{eq:eq_log}). We notice that only $EP_{LMD,t}$ retains its significance suggesting that LMD contains terms,  not included in EL, that are relevant in explaining the occurrence of recessions. These results are available upon request to the authors.}.  Overall, these results show that EL improves relative to LMD, SSW, and REN by providing more accurate sentiment scores and also by providing better coverage of terms that are relevant for forecasting recessions\footnote{In the online Appendix we perform this analysis dynamically by estimating model in Equation ~(\ref{eq:eq_log}) in expanding windows of data starting in January 2000. This exercise allows us to assess the importance of a more accurate sentiment score for the terms in disagreement and differences in word coverage of alternative lexicons with respect to $EL$ through time and particular periods such as recessions.
}.

\begin{table} \centering 
\begin{tabular}{@{\extracolsep{25pt}}lccc} 
\\[-1.8ex]\hline \hline \\[-1.8ex] 
\\[-1.8ex] 
& \multicolumn{3}{c}{$RECESSION_{t+3}$} \\ \cline{2-4}\\[-1.8ex]
 & LMD & SSW & REN \\ \hline \\[-1.8ex] 
  $EP_{l,t}$ & 0.928$^{***}$ & 0.435$^{***}$ & 0.582$^{***}$  \\ 
  & (0.161) & (0.142) & (0.168)  \\  
 $\Delta EP_{disagree,t}$ & 7.802$^{**}$ & 6.406$^{***}$ & 0.514  \\ 
  & (3.130) & (2.074) & (1.757) \\ 
 $\Delta EP_{onlyEL,t}$ & 1.060$^{***}$ & 2.361$^{*}$ & 0.939$^{***}$  \\ 
  & (0.314) & (1.246) &  (0.308)  \\ 
\hline \\[-1.8ex] 
Observations & 491 & 491 & 491 \\ 
Log Likelihood & $-$130.488 & $-$139.119 & $-$142.917 \\ 
Akaike Inf. Crit. & 272.976 & 290.237 & 297.834 \\ 
\hline \hline \\[-1.8ex] 
\end{tabular} 
  \caption{Estimation results for the $\gamma_i$ (with $i \in \{l, d, o\}$) parameters in Equation~(5). Results for $Spread_t$ and $ADS_t$ are omitted to save space. Standard errors are reported in parenthesis. $^{***}$ denotes significance at  1\%, $^{**}$ at 5\%,  and $^{*}$  at 10\% significance. }  
 \label{tab:tab1}
\end{table}

\color{black}

\section{Concluding remarks\label{conclusions}}

In this paper we develop a lexicon that is designed for applications of sentiment analysis to textual data in economics, such as speeches of governors or economic news. This is a tool very much in need as the evidence from several papers indicates that general purpose lexicons, or even lexicons developed in accounting and finance, do not provide the appropriate set of words and sentiment classification for economic applications. We build the lexicon from the ground up, by analyzing an extensive corpus of economic documents that provide a set of words that are typically used when discussing economic concepts. A unique feature of our dictionary is that we obtain the sentiment score of each term based on the interpretation of human annotators. We believe this aspect sets our Economic Lexicon (EL) apart from other dictionaries proposed in literature that derive the sentiment using a model-based approach.

In several applications we show that a measure of economic pessimism based on the EL seems to accurately capture the behavior of economic variables, such as measures of macroeconomic and financial uncertainty, consumer sentiment, and in predicting recessions. In addition, we show that the sentiment indicator constructed using the EL is, in most situations, more accurate relative to the pessimism based on the three alternative dictionaries. We find that this result is driven by two main factors. One is that the EL covers a wider set of terms that are typically used in economic text. The second element is that the EL sentiment score seems to reflect more precisely the economic interpretation of the terms. In both of these dimensions the EL seems to perform better relative to the alternative lexicons, although there is still room for improvement. One direction is to further extend the coverage of the lexicon, in particular to include more financial terms as in \cite{Loughran_McDonald_2011} and \cite{correa2021sentiment}. This could be relevant, in particular, when episodes of financial instability might trigger an economic slowdown. In addition, more extensive evaluation of the sentiment scores is needed, for example, by comparing the sentiment score for a whole sentence based on the EL with the score of a human annotator. We plan to develop these aspects in future work.

\section*{Acknowledgements}
The views expressed are purely those of the authors and may not in any circumstance be regarded as stating an official position of the European Commission.
We are greatly indebted to the Centre for Advanced Studies at the Joint Research Centre of the European Commission for the support, encouragement, and stimulating environment while working on the {\it bigNOMICS} project. We thank the participants of the 2022 Computational and Financial Econometrics conference and the 2023 Symposium of the Society of Nonlinear Dynamics and Econometrics for their insightful comments.

\section*{Data availability}
The data for \cite{barbaglia2024ECIN} that support the findings of this study are available in openICPSR at \url{https://doi.org/10.3886/E208164V3}, reference number \textit{openicpsr-208164}.

\newpage 

\setstretch{1.5}

\bibliographystyle{chicago}
\bibliography{ref}

\end{document}